\documentclass[10pt,twocolumn,letterpaper]{article}

\usepackage{cvpr}
\usepackage{times}
\usepackage{epsfig}
\usepackage{graphicx}
\usepackage{amsmath}
\usepackage{amssymb}
\usepackage{booktabs}
\usepackage{algorithm}
\usepackage{algorithmic}
\usepackage{tikz}
\usepackage{graphicx}

\title{Developing an Artificial Intelligence Tool for Personalized Breast Cancer Treatment Plans based on the NCCN Guidelines}

\author{Abdul M. Mohammed\thanks{abdul.mohammed@mercurial-ai.com} \quad 
Iqtidar Mansoor\thanks{iqtidar.mansoor@mercurial-ai.com} \quad 
Sarah Blythe\thanks{sarah.blythe@mercurial-ai.com} \quad 
Dennis Trujillo\thanks{dennis.trujillo@mercurial-ai.com}\\
\\ 
Mercurial AI Inc
}

\begin{document}

\maketitle

\begin{abstract}
Cancer treatments require personalized approaches based on a patient's clinical condition, medical history, and evidence-based guidelines. The National Comprehensive Cancer Network (NCCN) provides frequently updated, complex guidelines through visuals like flowcharts and diagrams, which can be time consuming for oncologists to stay current with treatment protocols. This study presents an AI (Artificial Intelligence)-driven methodology to accurately automate treatment regimens following NCCN guidelines for breast cancer patients.

We proposed two AI-driven methods: Agentic-RAG (Retrieval-Augmented Generation) and Graph-RAG. Agentic-RAG used a three-step Large Language Model (LLM) process to select clinical titles from NCCN guidelines, retrieve matching JSON content, and iteratively refine recommendations based on insufficiency checks. Graph-RAG followed a Microsoft-developed framework with proprietary prompts, where JSON data was converted to text via an LLM, summarized, and mapped into graph structures representing key treatment relationships. Final recommendations were generated by querying relevant graph summaries. Both were evaluated using a set of patient descriptions, each with four associated questions.

As shown in Table 1, Agentic RAG achieved a 100\% adherence (24/24) with no hallucinations or incorrect treatments. Graph-RAG had 95.8\% adherence (23/24) with one incorrect treatment and no hallucinations. Chat GPT-4 showed 91.6\% adherence (22/24) with two wrong treatments and no hallucinations. Both Agentic RAG and Graph-RAG provided detailed treatment recommendations with accurate references to relevant NCCN document page numbers.
\end{abstract}

\section{Introduction}
Cancer remains one of the most significant global health challenges of the 21st century. According to the latest GLOBOCAN estimates, approximately 20.1 million new cancer cases and 9.9 cancer deaths occurred worldwide in 2023 [1]. This number is projected to rise to 28.4 million cases by 2040, due to various factors such as population growth, aging, and changes in the prevalence and distribution of risk factors [2].

Advances in cancer care have led to the development of highly personalized treatment regimens tailored to individual patients based on tumor characteristics, molecular profiling, and clinical stage. However, as these treatment regimens grow more sophisticated, clinicians face the overwhelming task of integrating vast amounts of data from clinical trials, guidelines, and emerging therapies into their decision-making processes.

The NCCN guidelines are structured around clinical flowcharts and diagrams that present diagnostic and treatment pathways for specific cancer types, including but not limited to breast, lung, colorectal, and prostate cancers. These guidelines are updated multiple times per year, making it challenging for Oncologists and other clinicians to stay up to date with the latest treatment protocols.

\subsection{Related Work}
The rapid evolution of Artificial Intelligence (AI) and Natural Language Processing (NLP) technologies, especially Generative AI, has opened new avenues for enhancing clinical decision support systems in Oncology [4, 5]. Several Large Language Models (LLMs) have demonstrated remarkable capabilities in understanding and generating human-like text, making them potentially powerful tools for interpreting complex medical guidelines.

However, the direct application of LLMs to clinical decision-making is limited by their tendency to hallucinate or provide outdated information [9]. This limitation is particularly critical in Oncology where treatment recommendations can change rapidly based on the latest research and development in the field.

\section{Methodology}
\subsection{Data Preprocessing}
The NCCN guidelines, initially in PDF format, consist of medical flowcharts, treatment algorithms, and detailed cancer management guidelines for various cancer conditions and patient's characteristics. Since much of the information in these documents was in the form of visuals such as flowcharts and tables, emphasis was given to the pages containing these visuals. These pages were converted into JSON format, with each JSON object capturing the information from each page of the NCCN documents.

\subsection{Agentic RAG (Retrieval-Augmented Generation)}
The agentic-RAG methodology implemented here consists of four key components:

\subsubsection{Title Selection}
The first step involves selecting relevant titles from the available list of clinical guideline titles using GPT-4o with a prompt specifically designed to analyze the patient's description and question.

\subsubsection{JSON Retrieval}
Once titles are selected, the corresponding JSON objects are retrieved from the dataset, representing information from the respective pages of NCCN documents.

\subsubsection{Treatment Recommendation Generation}
The treatment generation phase uses a second LLM (o1-preview) with a prompt tailored to produce detailed and accurate treatment plans following a structured template.

\subsubsection{Insufficiency Check}
To validate the comprehensiveness of the generated recommendations, a third LLM call evaluates the sufficiency of the output using a detailed checklist of care aspects.

\begin{figure*}[t]
    \centering
    \includegraphics[width=\columnwidth]{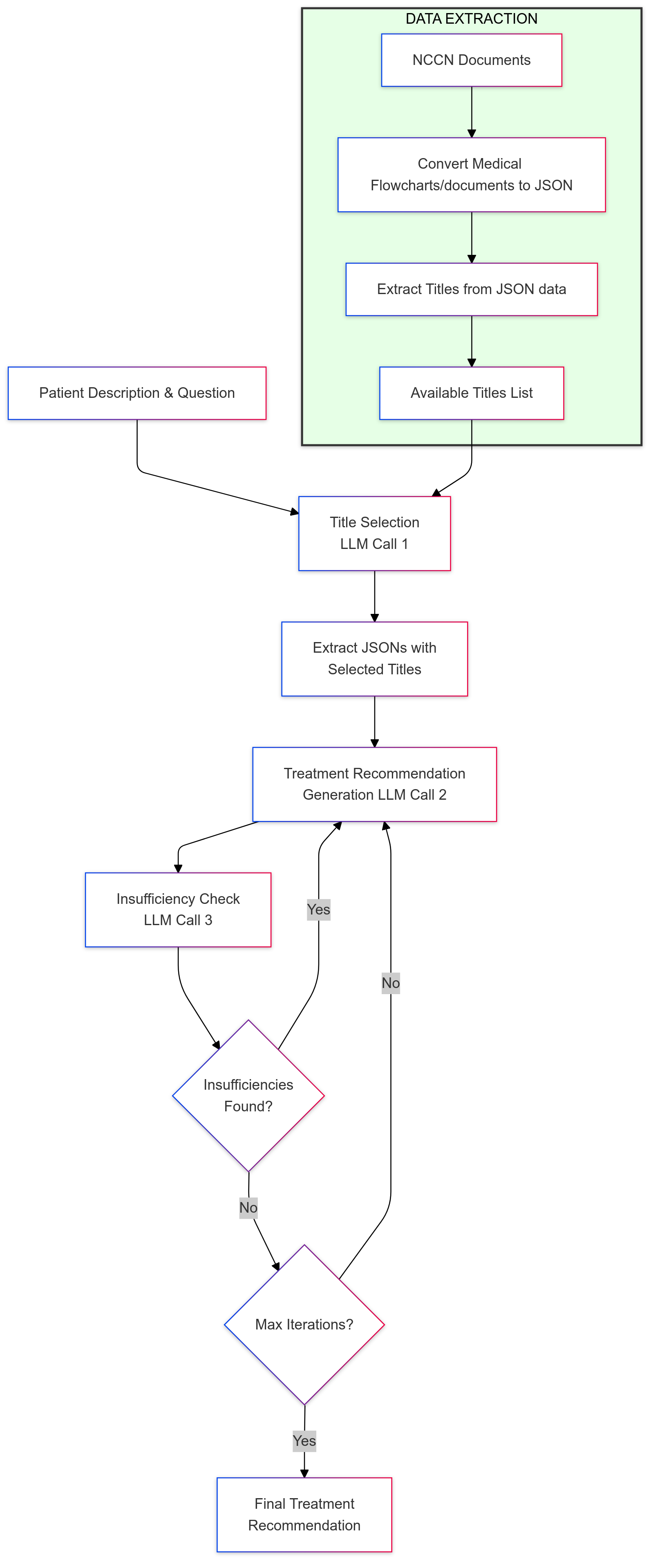}
    \caption{Agentic RAG-based Cancer Treatment Recommendation System using NCCN Data. The system incorporates three key LLM calls for title selection, recommendation generation, and insufficiency checking, with an iterative refinement process to ensure comprehensive and accurate recommendations.}
    \label{fig:agentic-rag}
\end{figure*}

\subsection{Graph-RAG}
The Graph-RAG approach consists of the following steps:

\begin{enumerate}
\item NCCN JSONs to Text Chunks
\item Text Chunks to Medical Entities and Relationships
\item Medical Entities and Relationships to Graph Element Summaries
\item Element Summaries to Graph Communities
\item Graph Communities to Community Summaries
\item Final Treatment Recommendation Generation
\end{enumerate}

\begin{figure*}[t]
    \centering
    \includegraphics[width=\columnwidth]{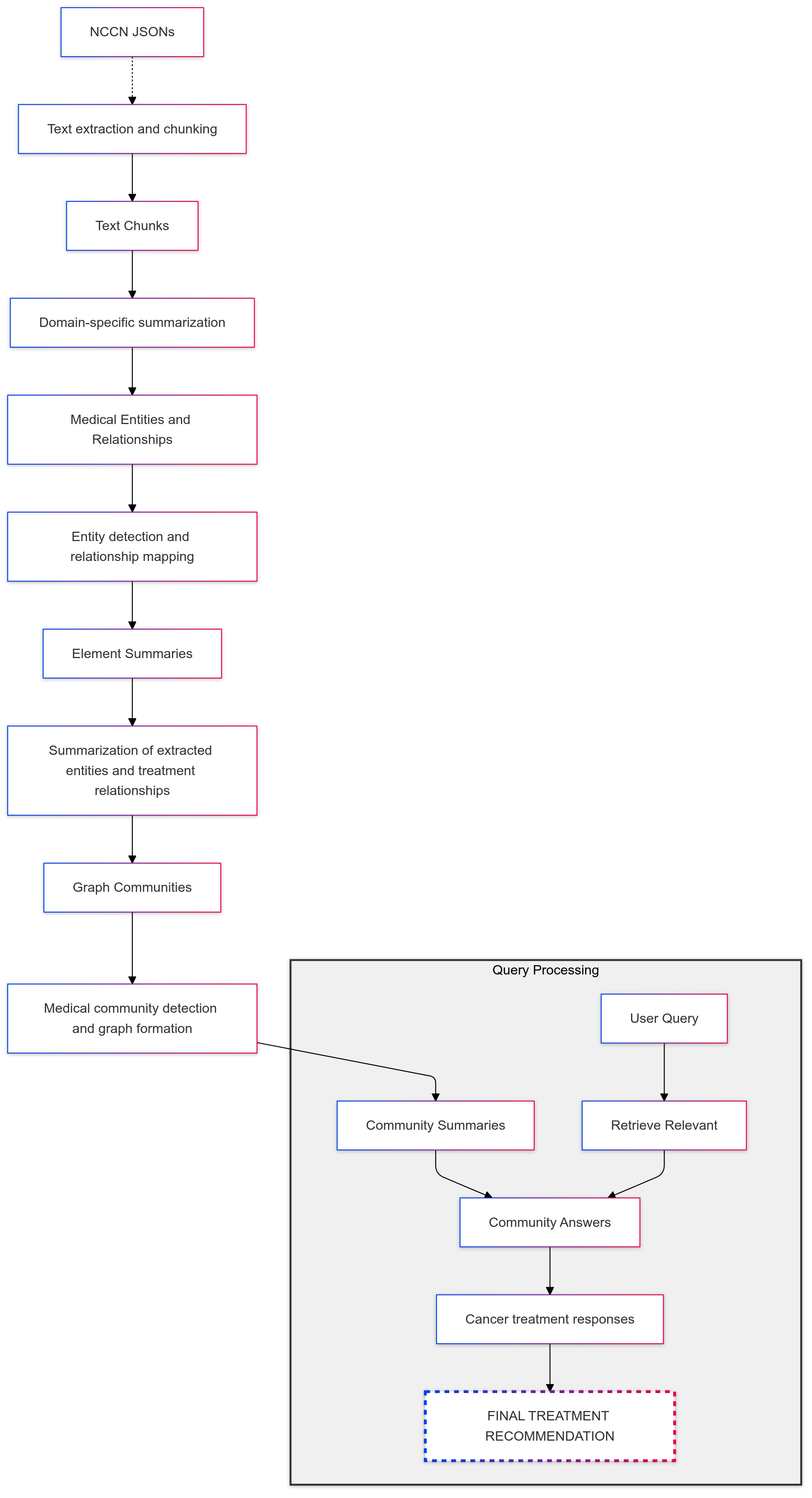}
    \caption{GraphRAG-based Cancer Treatment Recommendation System using NCCN Data. The system processes NCCN documents through multiple stages of text extraction, entity recognition, and graph formation before generating treatment recommendations based on user queries.}
    \label{fig:graph-rag}
\end{figure*}

\section{Experimental Setup}
\subsection{Patient Descriptions and Query Variations}
Sixteen unique patient descriptions were initially curated to represent a broad range of clinical scenarios involving breast cancer management. For each patient description, four distinct question variations were tested:

\begin{enumerate}
\item "What are the recommended treatments for this patient?"
\item "How should this patient be treated?"
\item "Provide a detailed treatment recommendation for this patient."
\item "What treatments align with NCCN guidelines for this case?"
\end{enumerate}

\subsection{Evaluation Criteria}

Each system's responses were manually evaluated by a board-certified physician using the following metrics:

\begin{itemize}
\item Number of Treatments: The number of treatment recommendations provided by the AI (0 for none, 1 for one treatment recommended, and 2 for multiple treatments recommended).
\item Hallucinations: Whether the system introduced treatments not found in the NCCN guidelines (0 for none, 1 for at least one).
\item Adherence Rate: The proportion of treatment recommendations that strictly adhered to the NCCN guidelines.
\item Error Types:
    \begin{itemize}
    \item Missing Treatments: Relevant treatments omitted in the recommendations.
    \item Unnecessary Treatments: Irrelevant or superfluous treatments included in the response.
    \item Wrong Treatments: Recommendations contradicting the NCCN guidelines.
    \item Order Errors: Misalignment in the logical or chronological order of recommendations.
    \end{itemize}
\end{itemize}

The performance of the three systems—ChatGPT-4, Graph-RAG (LLM V1), and Agentic-RAG (GPT-4o)—was evaluated based on adherence to NCCN guidelines, hallucination rates, and treatment-related errors. The evaluation framework followed the benchmarking methodology described by Chen et al. [24]

\section{Results and Discussion}
\begin{table}
\begin{center}
\begin{tabular}{lccc}
\toprule
Metric & ChatGPT-4 & LLM V1 & Agentic RAG \\
\midrule
Hallucinations & 0 & 0 & 0 \\
Adherence rate & 94\% & 92\% & 100\% \\
Wrong order & 0 & 0 & 0 \\
Unnecessary & 1 & 0 & 0 \\
Missing treatments & 2 & 4 & 0 \\
Wrong treatments & 0 & 0 & 0 \\
\bottomrule
\end{tabular}
\end{center}
\caption{Performance comparison of systems in adherence to treatment guidelines.}
\label{table:results}
\end{table}

\subsection{System Performance}
The performance of three systems—ChatGPT-4, Graph-RAG, and Agentic-RAG—was evaluated based on
adherence to NCCN guidelines, hallucination rates, and treatment-related errors. The results highlight the
strengths and limitations of each system, as summarized in the table 1.

Agentic-RAG achieved a perfect adherence rate of 100\%, correctly identifying all 50 treatment
recommendations in alignment with NCCN guidelines. In addition to providing precise and complete
recommendations, Agentic-RAG included detailed explanations for each treatment, along with references to the
specific NCCN documents and page numbers from which the information was extracted. This ability to
reference the exact source for each treatment ensures transparency and reliability, making it particularly suited
for clinical applications.

Graph-RAG, while achieving a slightly lower adherence rate of 92\%, similarly provided detailed treatment
recommendations with NCCN document and page references. Its graph-based approach facilitated the
extraction of structured relationships between clinical entities and treatments, enabling detailed and
NCCN-aligned outputs. However, occasional omissions of nuanced or overlapping guideline information led to
the system missing four treatments.

ChatGPT-4, as a general-purpose LLM, achieved an adherence rate of 94\%, missing two treatments and
including one unnecessary recommendation. While its outputs were accurate and avoided hallucinations, the
responses lacked the depth and clinical alignment achieved by the proposed systems. Notably, ChatGPT-4 did
not provide NCCN document references, which limits its applicability in clinical settings where transparency
and source validation are critical.

\subsection{Key Findings}
None of the systems hallucinated treatments, demonstrating strong baseline reliability. Similarly, no logical
sequencing errors or incorrect treatments were observed in all systems, reflecting their adherence to the logical flow of treatments once the correct data was recovered. However, the proposed systems, Agentic-RAG and Graph-RAG, excelled in providing detailed and clinically aligned treatment recommendations. They uniquely offered references to exact NCCN documents and page numbers for each treatment, enhancing their reliability and usability in clinical workflows.

Agentic-RAG's structured methodology, including iterative sufficiency checks, ensured complete alignment
with NCCN guidelines, achieving a perfect adherence rate and providing the most comprehensive outputs.
Graph-RAG also provided detailed and clinically aligned recommendations, although its reliance on graph-based
retrieval occasionally resulted in missed treatments. ChatGPT-4, while performing well as a baseline, lacked the
depth, specificity, and reference transparency of the proposed systems.

\section{Conclusion}
In this study, we developed and evaluated two new systems, AgGentic-RAG and Graph-RAG, to generate cancer treatment recommendations based on NCCN guidelines. These systems were designed to address the limitations of general-purpose LLMs by incorporating structured retrieval methodologies and utilizing domain-specific optimizations.

Agentic-RAG achieved perfect performance across all evaluated metrics, providing detailed NCCN-aligned treatment recommendations with transparent references to specific NCCN documents and page numbers. Its iterative and LLM-guided approach not only retrieves accurate and comprehensive information, but also delivers
outputs that are highly reliable and actionable in clinical workflows. This represents a significant advance in the use of AI for evidence-based cancer treatment recommendations.

Graph-RAG also provided detailed and referenced recommendations, though it was occasionally limited by its reliance on graph-based retrieval to capture nuanced or overlapping guideline content. Despite these limitations, it significantly outperformed the baseline model in adherence to guidelines and transparency.

In contrast, ChatGPT-4, while demonstrating strong baseline capabilities, lacked the structured mechanisms
required for complete and clinically transparent outputs. This limitation reduces its applicability in high-stakes
medical settings where accuracy and source verification are paramount.

\subsection{Clinical Impact and Future Directions}
The ability of both proposed systems to integrate detailed treatment recommendations with clear NCCN
references highlights their potential to transform clinical practice. By bridging the gap between comprehensive
clinical guidelines and personalized patient care, these systems pave the way for AI-driven decision support
tools that enhance both accuracy and trust in medical applications.

Future work will involve broadening the testing of Agentic-RAG and Graph-RAG by evaluating them on a
more extensive set of patient descriptions across various types and stages of cancer. This expanded dataset will
help assess the systems' performance in diverse clinical scenarios. Additionally, involving more oncologists in
the evaluation process will provide critical clinical insights and feedback. Their expertise will be invaluable in
assessing the practicality, relevance, and acceptance of the systems' recommendations in real-world settings.

Moreover, integrating these systems with electronic health records (EHRs) and real-time decision-making
workflows could further enhance their clinical impact. Ensuring interoperability with existing clinical systems
and compliance with healthcare regulations will be essential steps toward successful implementation.

By enhancing the accuracy, transparency, and efficiency of treatment planning, Agentic-RAG and Graph-RAG
offer a promising step toward integrating AI-driven tools in clinical oncology. Their ability to provide
transparent, NCCN-aligned recommendations with detailed references supports clinicians in delivering
personalized and guideline-compliant cancer care, ultimately aiming to improve patient outcomes.


\begin{thebibliography}{24}
\bibitem{ref1} Sung H, Ferlay J, Siegel RL, Laversanne M, Soerjomataram I, Jemal A, et al. Global Cancer Statistics 2024: GLOBOCAN Estimates of Incidence and Mortality Worldwide for 36 Cancers in 185 Countries. CA Cancer J Clin. 2024;74(1):29-52.

\bibitem{ref2} Cao B, Soerjomataram I, Bray F, Fidler-Benaoudia MM. The global burden of cancer attributable to risk factors, 2010–2019: a systematic analysis for the Global Burden of Disease Study 2019. Lancet. 2023;401(10376):563-585.

\bibitem{ref3} National Comprehensive Cancer Network. NCCN Guidelines for Breast Cancer. Available from: https://www.nccn.org/professionals/physician\_gls/default.aspx\#breast

\bibitem{ref4} Yu KH, Beam AL, Kohane IS. Artificial intelligence in healthcare. Nat Biomed Eng. 2018 Oct;2(10):719-731.

\bibitem{ref5} Duwe G, Mercier D, Wiesmann C, Kauth V, Moench K, Junker M, et al. Challenges and perspectives in use of artificial intelligence to support treatment recommendations in clinical oncology. Cancer Medicine. 13.

\bibitem{ref6} Brown TB, Mann B, Ryder N, Subbiah M, Kaplan J, Dhariwal P, et al. Language Models are Few-Shot Learners. arXiv preprint arXiv:2005.14165.

\bibitem{ref7} Schulte B. Capacity of ChatGPT to Identify Guideline-Based Treatments for Advanced Solid Tumors. Cureus. 2023 Apr 21;15(4):e37938.

\bibitem{ref8} Piazza D, Martorana F, Curaba A, Sambataro D, Valerio MR, Firenze A, et al. The Consistency and Quality of ChatGPT Responses Compared to Clinical Guidelines for Ovarian Cancer: A Delphi Approach. Current Oncology. 2024;31(5):2796-2804.

\bibitem{ref9} Koopman B, Zuccon G. Artificial intelligence in medical decision making: challenges for research and practice. Med J Aust. 2023;219(6):263-265.

\bibitem{ref10} Kuck G. Tim Berners-Lee's Semantic Web. South African Journal of Information Management. 2004;6.

\bibitem{ref11} Lewis P, Perez E, Piktus A, Petroni F, Karpukhin V, Goyal N, et al. Retrieval-Augmented Generation for Knowledge-Intensive NLP Tasks. Adv Neural Inf Process Syst. 2023;33:9459-9474.

\bibitem{ref12} Procko T, Ochoa O. Graph Retrieval-Augmented Generation for Large Language Models: A Survey. Available at SSRN: https://ssrn.com/abstract=4895062

\bibitem{ref13} Finsås M, Maksim J. Optimizing RAG Systems for Technical Support with LLM-based Relevance Feedback and Multi-Agent Patterns. Master's thesis, Norwegian University of Science and Technology (NTNU).

\bibitem{ref14} Das R, Maheswari K, Siddiqui S, Arora N, Paul A, Nanshi J, et al. Improved precision oncology question-answering using agentic LLM. medRxiv.

\bibitem{ref15} Edge D, Trinh H, Cheng N, Bradley J, Chao A, Mody A, et al. From Local to Global: A Graph RAG Approach to Query-Focused Summarization. Microsoft Research.

\bibitem{ref16} Wang J, Shi E, Yu S, Wu Z, Ma C, Dai H, et al. Prompt Engineering for Healthcare: Methodologies and Applications. Journal of Biomedical Informatics. 2021;14(8):1.

\bibitem{ref17} Sahoo P, Singh AK, Saha S, Jain V, Mondal S, Chadha A. A Systematic Survey of Prompt Engineering in Large Language Models: Techniques and Applications. arXiv:2402.07927.

\bibitem{ref18} Chen B, Zhang Z, Langrené N, Zhu S. Unleashing the potential of prompt engineering in Large Language Models: a comprehensive review. arXiv:2310.14735.

\bibitem{ref19} National Comprehensive Cancer Network. NCCN Clinical Practice Guidelines in Oncology: Multiple Myeloma, Version 2.2024. Journal of the National Comprehensive Cancer Network. 2023;21(12):1281-1320.

\bibitem{ref20} National Comprehensive Cancer Network. Management of Immunotherapy-Related Toxicities, Version 1.2024. Journal of the National Comprehensive Cancer Network. 2023;21(5.5):556-567.

\bibitem{ref21} National Comprehensive Cancer Network. Breast Cancer, Version 3.2024. Journal of the National Comprehensive Cancer Network. 2024;22(5):331-339.

\bibitem{ref22} OpenAI. Introduction to GPT-4o. OpenAI Cookbook.

\bibitem{ref23} OpenAI. O1-small Model. OpenAI.

\bibitem{ref24} Chen S, Kann BH, Foote MB, et al. Use of Artificial Intelligence Chatbots for Cancer Treatment Information. JAMA Oncol. 2023;9(10):1459-1462.
\end{thebibliography}
\end{document}